%% file: main.tex
\documentclass[submission,copyright,creativecommons]{eptcs}
\providecommand{\event}{WoC 2015} 
\usepackage{breakurl}             

\usepackage{amsthm}
\usepackage{amsmath,amssymb}
\usepackage{bussproofs}

\usepackage{stmaryrd} 

\usepackage{listings}

\lstdefinestyle{ocaml}{
language=[Objective]Caml,
basicstyle={\small\usefont{T1}{pcr}{m}{n}},   
keywordstyle={\small\usefont{T1}{pcr}{b}{n}}, 
otherkeywords={\%,->},
keywords={[3]{@,@@,\%,->}},
literate={
    {->}{$\to$}1
    {'a}{$\alpha$}1
    {'b}{$\beta$}1
    {'c}{$\gamma$}1
    {'t}{$\tau$}1
    {'s}{$\sigma$}1
},
commentstyle={\itshape},
mathescape=true,
showtabs=false,
showspaces=false,
showstringspaces=false,
keepspaces=true,
tabsize=4,
breaklines=true,
xrightmargin=0em,
xleftmargin=3em,
columns=[l]{fullflexible},
}

\newcommand{\jump}[1]{\ensuremath{\llbracket #1 \rrbracket} }
\newcommand{\jumpone}[1]{\ensuremath{\llbracket #1 \rrbracket}^{1pass}}
\newcommand{\jumptwo}[1]{\ensuremath{\llbracket #1 \rrbracket}^{opt}}
\newcommand{\prompt}[1]{#1 {\ \text{pr}}}
\newcommand{\letexp}[2]{\text{let}\ #1 \ \text{in}\ #2}

\newcommand\shift{\mathcal{S}}
\newcommand{\resetp}[2]{\langle #2 \rangle_{#1}}
\newcommand\newp{\mathcal{P}}
\newcommand{\throw}[2]{\text{throw}(#1,#2)}

\newcommand{\resetpu}[2]{\underline{\langle} #2 \underline{\rangle_{#1}}}

\newcommand{\letexpu}[2]{\underline{\text{let}}\ #1 \ \underline{\text{in}}\ #2}

\newcommand\lamATM{\lambda^{\textrm{ATM}}}
\newcommand\hole{[\ ]}
\newcommand\mpsr{\lambda^{\textrm{mpsr}}}

\newcommand{\puretr}[1]{\ensuremath{\llparenthesis #1 \rrparenthesis} }
\newcommand{\puretrone}[1]{\ensuremath{\llparenthesis #1 \rrparenthesis}^{1pass}}
\newcommand{\puretrtwo}[1]{\ensuremath{\llparenthesis #1 \rrparenthesis}^{opt}}

\newcommand{\slam}[2]{\overline{\lambda}{#1}.{#2}}
\newcommand{\slamtwo}[3]{\slam{{#1}{#2}}{#3}}
\newcommand{\sapp}[2]{{#1}\overline{@}{#2}}
\newcommand{\sapptwo}[3]{\sapp{\sapp{#1}{#2}}{#3}}

\newcommand{\dlam}[2]{\underline{\lambda}{#1}.{#2}}
\newcommand{\dlamtwo}[3]{\dlam{{#1}{#2}}{#3}}
\newcommand{\dapp}[2]{{#1}\underline{@}{#2}}
\newcommand{\dapptwo}[3]{\dapp{\dapp{#1}{#2}}{#3}}
\newcommand{\dnewp}[2]{\underline{\newp}{#1}.{#2}}
\newcommand{\dnewptwo}[3]{\dnewp{#1}{\dnewp{#2}{#3}}}  
\newcommand{\dshift}[3]{\underline{\shift_{#1}}{#2}.{#3}}
\newcommand{\Int}{\texttt{int}}

\newtheorem{theo}{Theorem}

\title{Answer-Type Modification without Tears:\\
Prompt-Passing Style Translation\\
for Typed Delimited-Control Operators}
\author{Ikuo Kobori 
\institute{University of Tsukuba, Japan}
\email{ikuo@logic.cs.tsukuba.ac.jp}
\and
Yukiyoshi Kameyama
\institute{University of Tsukuba, Japan}
\email{kameyama@acm.org}
\and
Oleg Kiselyov
\institute{Tohoku University, Japan}
\email{oleg@okmij.org}
}
\def\titlerunning{ATM without Tears}
\def\authorrunning{I. Kobori, Y. Kameyama \& O. Kiselyov}
\begin{document}
\maketitle

\begin{abstract}
The salient feature of delimited-control operators is their ability
to \textit{modify} answer types during computation.
The feature, answer-type modification (ATM for short), allows one
to express various interesting programs such as typed printf
compactly and nicely, 
while it makes it difficult to embed these operators
in standard functional languages.

In this paper, we present a typed translation of delimited-control
operators shift and reset with ATM into a familiar language 
with multi-prompt shift and reset without ATM, which
lets us use ATM in standard languages without modifying 
the type system.
Our translation generalizes Kiselyov's direct-style
implementation of typed printf, which uses two prompts to 
emulate the modification of answer types, and passes them 
during computation.
We prove that our translation preserves typing.
As the naive prompt-passing style translation generates and
passes many prompts even for pure terms, we show an optimized
translation that generate prompts only when needed, which is 
also type-preserving.
Finally, we give
an implementation in the tagless-final style which respects typing 
by construction.
\end{abstract}

\input{introduction}

\input{atm}

\input{simulation_atm}

\input{calculi}

\input{translation}

\input{opt}

\input{tagless_final}

\input{example}

\input{conclusion}

\paragraph{Acknowledgments:}
We are grateful to the anonymous reviewers and
the audience at WoC'15 for their constructive comments, 
and the editors of the post-proceedings, Olivier Danvy and
Ugo de'Liguoro for suggestions and advice.
The authors are supported in part by JSPS Grant-in-Aid for
Scientific Research (25280020, 25540023).

\bibliographystyle{eptcs}
\bibliography{ref}


\end{document}

%% file: introduction.tex

\section{Introduction} \label{sec:intro}

Delimited continuations are parts of a continuation, the rest of computation, and
delimited-control operators provide programmers
a means to access the current delimited continuations.
Since the delimited-control operators control/prompt and shift/reset 
have been proposed around 1990 \cite{Felleisen1988,Danvy1990}, 
many researchers have been studying them intensively, 
to find interesting theory and application in program transformation,
partial evaluation, code generation, 
and computational linguistics.
Today, we see their implementations in many programming languages such as
Scheme, Racket, SML, OCaml, Haskell, and Scala.

Yet, there still exists a big gap between theory and practice
if we work in typed languages.
Theoretically, the salient feature of delimited-control operators
is their ability to modify answer types.
The term \verb|reset (3 + shift k -> k)| looks as if it has type \verb|int|, but
the result of this computation is a continuation \verb|fun x -> reset (3 + x)|
whose type is \verb|int -> int|, which means that
the initial answer type has been modified during the computation of the
shift term. While this feature, called Answer-Type Modification, 
allows one to express surprisingly interesting programs such as 
typed printf, it is the source of the problem that 
we cannot easily embed the delimited-control operators
in standard languages.
We can hardly expect that the whole type system 
of a full-fledged language would be modified in such a way.
With a few exceptions of Scala \cite{Rompf2009} and OchaCaml
\cite{Masuko2011}, we cannot directly express the beautiful examples
with ATM as programs in standard languages.

We address this problem, and present a solution for it.
Namely, we give a translation from a calculus with
ATM shift and reset into a calculus with multi-prompt shift and
reset without ATM.  Our translation is a generalization of Kiselyov's
implementation~\cite{Kiselyov2007a} of typed printf using multi-prompt shift and reset
where he associates each answer type with a prompt (a tag)
for delimited-control operators.
Our translation also uses prompts to simulate answer types,
and the term obtained by our translation 
dynamically generates and passes two prompts during computation,
thus we call it Prompt-Passing Style (PPS), after the well known 
Continuation-Passing Style.

We introduce a PPS translation from
a calculus with ATM to a calculus without, and prove that it 
is type-preserving.  We also give an implementation based on our 
translation in the tagless-final style~\cite{Carette2009,Kiselyov:2010:TTF}, which allows us
to embed a domain-specific language while preserving types by construction.

The PPS translation differs from 
the definitional CPS translation for shift and reset
\cite{Danvy1990} in that the generated term by our PPS translation 
are in direct style, while the generated terms by the CPS translation
are in continuation-passing style, which makes the size of terms bigger,
and may affect performance.
In order to show this aspect better, we refine the naive PPS translation
to obtain an optimized PPS translation, where
prompts are generated and passed only when needed.
It is based on the idea of one-pass CPS translation 
as well as an ad hoc optimization for prompts.
The optimized translation is also type preserving and has been implemented
in the tagless-final style.  
We show by examples that the optimized PPS translation 
generates much smaller terms than the naive PPS translation
and the CPS translation.

The rest of this paper is organized as follows: 
Section \ref{sec:atm} explains 
delimited-control operators and answer-type modification 
by a simple example.
Section \ref{sec:simulation} informally
states how we simulate answer-type modification
using multi-prompt shift and reset, and 
Section \ref{sec:formal} gives a formal account to it including
formal properties.
Section \ref{sec:trans} describes the syntax-directed translation and
its property, and Section \ref{sec:opt} introduces an optimized translation
with examples.
Based on these theoretical developments, Section \ref{sec:tagless_final} 
gives a tagless-final implementation of shift and reset with answer-type
modification as well as several programming examples.
Section \ref{sec:conclusion} gives related work and concluding remarks.

%% file: atm.tex

\section{Delimited-Control Operators and Answer-type Modification} \label{sec:atm}

We introduce a simple example which uses delimited-control operators shift 
and reset where the answer types are modified through computation.

The following implementation of the \verb|append| function is taken from
Danvy's paper~\cite{Danvy:ILFL88}. This program uses \verb|shift|
operator instead of \verb|call/ct| operator.

\begin{lstlisting}[style=ocaml]
let rec append lst = match lst with
  | [] -> shift (fun k -> k)
  | x :: xs -> x :: append xs
in let append123 = 
  reset (append [1;2;3])
in
  append123 [4;5;6]
\end{lstlisting}

The function \verb|append| takes a value of type \verb|int list| as its
input, and traverses the list.
When it reaches the end of the list, it captures the continuation
(\verb|fun ys -> reset 1 :: 2 :: 3 :: ys| in the functional form)
up to the nearest reset, and returns the continuation as its result.
We then apply it to the list \verb|[4;5;6]| to obtain 
\verb|[1;2;3;4;5;6]|, and it is easy to see
that the function deserves its name.

Let us check the type of \verb|append|.
At the beginning, the return type of \verb|append| (called its answer type)
is \verb|int list|, 
since in the second branch of the case analysis,
it returns \verb|x :: append xs|.
However, the final result is a function from list to list, which is
different from our initial guess.
The answer type has been modified during the execution of the program.

Since its discovery, this feature has been used in
many interesting examples with shift and reset, from 
typed printf to suspended computations, to coroutines, and even
to computational linguistics. Nowadays,
it is considered as one of the most attractive features of shift and reset.

Although the feature, answer-type modification, is interesting and sometimes useful, 
it is very hard to
directly embed such control operators in conventional functional
programming languages such as OCaml, as it requires a big change
of the type system; a typing judgment in the form $\Gamma \vdash e:\tau$
must be changed to a more complex form 
$\Gamma \vdash e:\tau; \alpha, \beta$ where $\alpha$ and $\beta$
designate the answer types before and after the execution of $e$.
Although adjusting a type system in this way is 
straightforward in theory, it is rather difficult to modify 
existing implementations of type systems, and we therefore need a way to represent
the above features in terms of standard features and/or mild extensions of 
existing programming languages.

This paper addresses this problem, and proposes a way to translate
away the feature of ATM using multi-prompt control operators.

%% file: simulation_atm.tex
\section{Simulating ATM with Multi-prompt shift/reset} \label{sec:simulation}

In this section, we explain the basic ideas of our translation.
Kiselyov implemented typed printf in terms of shift and reset without ATM, 
and we have generalized it to a translation from arbitrary terms 
in the source language.

Consider a simple example with answer-type modification:
\ $\puretr{\resetp{}{5 + \shift k.k}}$
in which $\shift$ is the delimited-control operator shift, and
$\resetp{}{\cdots}$ is reset.
Its answer type changes through computation, as its initial answer type is
\verb|int| while its final answer type is \verb|int->int|.

Let us translate the example 
where
$\puretr{e}$ and $\jump{e}$ denote the results of the translations of the term $e$.
(The precise definition of the translations are given later.)

We begin with the translation of a reset expression:
\[
\puretr{\resetp{}{e}} = \newp p.\newp q. \resetp{p}{\letexp{y=\jump{e}p q}{\shift_q z.~y}} 
\]
where the primitive $\newp p$ creates a new prompt and binds the variable $p$ to it.
For brevity, the variable $p$ which stores a prompt may also be called a prompt.

The translated term, when it is executed,
first creates new prompts $p$ and $q$ and its body
$e$ is applied to the arguments $p$ and $q$. Its result is stored in $y$ and
then we execute $\shift_q z.y$, but there is no reset with the prompt $q$
around it.
Is it an error $?$  Actually, no.  As we will see in the definition below,
$\jump{e}$ is always
in the form $\lambda p.\lambda q. e'$ and during the computation of $e'$,
$\shift_p$ is \textit{always} invoked.
Hence $e'$ never returns normally, and 
the ``no-reset'' error does not happen.
Our invariants in the translation are that the 
first argument (the prompt $p$) corresponds to
the reset surrounding the expression being translated, and
the second argument (the prompt $q$) corresponds to 
the above (seemingly dangerous) shift.

From the viewpoint of typing, 
for each occurrence of answer-type modification 
from $\alpha$ to $\beta$,
we use two prompts to simulate the behavior.
The prompts $p$ and $q$ generated here correspond to the answer types 
$\beta$ and $\alpha$, respectively.

We translate the term $5$ to
$
\jump{5} = \lambda p.\lambda q.~\shift_p k.~\resetp{q}{k~5} 
$
and the term $\resetp{}{5}$ is translated (essentially) to:
\[
\newp p.\newp q.
\resetp{p}{\letexp{y=\shift_p k.\resetp{q}{k~5}}{\shift_q z.~y}}
\]
When we execute the result, $\shift_p$ captures its surrounding 
evaluation context $\resetp{p}{\letexp{y=\hole}{\shift_q z.~y}}$,
binds $k$ to its functional form
$\lambda x.\resetp{p}{\letexp{y=x}{\shift_q z.~y}}$,
and continues the evaluation of $\resetp{q}{k~5}$.
Then we get:
\[
\resetp{p}{\resetp{q}{\resetp{p}{\letexp{y=5}{\shift_q z.~y}}}}
\]
and when this $\shift_q$ is invoked, 
it is surrounded by a reset with the prompt $q$, and thus it is \textit{safe}.
The final result of this computation is $5$. 
In this case, since the execution of the term $5$ does not
modify the answer type, the prompts $p$ and $q$ passed to the term $\jump{5}$
correspond to the same answer type, but we will soon see an example
in which they correspond to different answer types.

A shift-expression is translated to:
\[
\jump{\shift k.e} = 
\lambda p. \lambda q. 
\shift_p k'. 
\letexp{k=(\lambda y.\resetp{q}{(\lambda \_.\Omega)(k' y)})}{\puretr{e}}
\]
As we have explained, $p$ is the prompt for the reset surrounding this
expression, hence $\shift_p$ in the translated term will capture a delimited continuation
up to the reset (which, in turn, corresponds to the nearest reset
in the source term).  However the delimited continuation contains
a dangerous shift 
at its top position, so we must somehow detoxify it.
For this purpose, we replace the captured continuation $k'$ by a 
function $\lambda y.\resetp{q}{(\lambda \_.\Omega)(k' y)}$
in which the calls to $k'$ is enclosed by a reset with the prompt $q$, and
the dangerous shift in $k'$ will be surrounded by it, 
sanitizing the dangerous behavior.

Let us consider the types of captured continuations in this translation. 
Suppose the term $\shift k.e$ modifies the answer type from $\alpha$ to $\beta$.
We use the prompts $p$ and $q$, whose answer types\footnote{We assume that,
our target language after the translation has multi-prompt shift and reset,
but no answer-type modification. Hence, each prompt has 
a unique answer type.}
are $\beta$ and $\alpha$, respectively.
In the source term, 
the continuation captured by shift (and then bound to $k$) 
has the type $\tau \to \alpha$.
In the translated term, 
the continuation bound to $k'$ 
has the type $\tau \to \beta$, since the continuation was captured
by a shift with the prompt $p$.
After some calculation, it can be inferred that 
the term $\lambda y.\resetp{q}{(\lambda \_.\Omega)(k' y)}$ has the
type $\tau \to \alpha$, hence we can substitute it for $k$.
\footnote{Here $\Omega$ is a term which has an arbitrary type. 
Such a term can be expressed,
as, for instance, $\newp p. \shift_p k. \lambda x.x$. Its operational behavior does not matter,
as it will be never executed.}

We show the mechanism for detoxifying a dangerous shift by executing
$\puretr{\resetp{}{5 + \shift k.k}}$, which is equivalent to:
\[
\newp p.\newp q. \resetp{p}{\letexp{y=
 \newp r.
 ((\shift_{r} k. \resetp{q}{k\, 5})
 +
 (\shift_{p} k'. 
   \letexp{k=\lambda u.\resetp{r}{(\lambda w. \Omega)(k'\, u)}}{k}))
 }{\shift_q z .y}}
\]
where the subterm starting with $\shift_r$ is the translation result of
$5$, and the one with $\shift_p$ is that of $\shift k.k$.
In general, each subterm may modify answer types. Hence, a term
$e_1 + e_2$ needs three prompts corresponding to the initial, final, and
intermediate answer types. The prompt $r$ generated here corresponds to the
intermediate answer type.

Evaluating this term in call-by-value, and right-to-left order 
(after generating all the prompts) leads to the term:
$
\resetp{p}{
\letexp{k=
\lambda u.\resetp{r}{(\lambda w. \Omega)(k'\, u)}}{k}
}
$
where $k'$ is  the delimited continuation 
$ \lambda x.
 \resetp{p}{\letexp{y=
 (\shift_{r} k. \resetp{q}{k\, 5})
 + x
 }{\shift_q z .y}} 
$. 
The result of this computation is
$\lambda u.\resetp{r}{(\lambda w. \Omega)(k'\, u)}$, 
which is essentially equivalent to $\lambda y.\resetp{}{5+y}$.
To see this, applying it to 9 yields:
\begin{align*}
 & (\lambda u.\resetp{r}{(\lambda w. \Omega)(
   (\lambda x.
 \resetp{p}{\letexp{y=
 (\shift_{r} k. \resetp{q}{k\, 5})
 + x
 }{\shift_q z .y}} )
   u)})\, 9 \\
 \leadsto^* &
 \resetp{r}{(\lambda w. \Omega)
 \resetp{p}{\letexp{y=
 (\shift_{r} k. \resetp{q}{k\, 5})
 + 9
 }{\shift_q z .y}}} \\
 \intertext{$\shift_r k. \resetp{q}{k\, 5}$ captures the context with
 the dangerous shift}
 \leadsto^* &
 \resetp{r}{\resetp{q}{
 (\lambda u.
 \resetp{r}{(\lambda w. \Omega)
 \resetp{p}{\letexp{y=u + 9}{\shift_q z .y}}})
 5}} \\
 \leadsto^* &
 \resetp{r}{\resetp{q}{
 \resetp{r}{(\lambda w. \Omega)
 \resetp{p}{\letexp{y=5 + 9}{\shift_q z .y}}}
 }} \\
 \leadsto^* & \resetp{r}{\resetp{q}{14}} \quad \text{which reduces to 14}.
\end{align*}

Thus, our translation uses two prompts to make connections to two answer
types, where prompts are generated dynamically.

%% file: calculi.tex

\section{Source and Target Calculi} \label{sec:formal}

In this section, we formally define our source and target calculi.

The source calculus is based on Asai and Kameyama's polymorphic extension of 
Danvy and Filinski's calculus for shift and reset, both of which
allow answer-type modification~\cite{Danvy1989,Asai2007}.
We slightly modified it here; (1) we removed 
fixpoint and conditionals (but they can be added easily), 
(2) we use value restriction for let-polymorphism 
while they used more relaxed condition,
and (3) 
we use Biernacka and Biernacki's simplification for the types of delimited continuations
~\cite{Biernacka2009}.

The syntax of 
values and terms of our source calculus $\lamATM$ is defined as follows:
 \begin{center}
 \begin{tabular}{rrrl}
 (values) & $v$& $::=$ & $x \mid c \mid \lambda x.e$ \\
 (terms) & $e$& $::=$ & $v \mid e_1\,e_2 \mid \letexp{x=v}{e} \mid \shift k.e
	 \mid \throw{k}{e}
	 \mid \resetp{}{e}$
 \end{tabular} 
 \end{center}
where $x$ is an ordinary variable,
$k$ is a continuation variable, and
$\throw{k}{e}$ is application for continuations,
which is syntactically different from ordinary application $e_1\,e_2$.
This distinction is technical and inessential for expressivity, 
as we can always convert a continuation $k$ to
a value $\lambda x.\throw{k}{x}$.
The variables $x$ and $k$, resp., are bound 
in the terms $\lambda x.e$ and $\shift k.e$, resp.,

Figure~\ref{fig:source-semantics} defines
call-by-value operational semantics to the language above, where
$\hole$ denotes the empty context and $E[e]$ denotes the usual 
hole-filling operation.
\begin{figure}[t]
\begin{tabular}{rccl}
 (evaluation contexts) & $E$ &$::=$&$ \hole \mid e\,E \mid E\,v \mid \resetp{}{E}$ \\
 (pure evaluation contexts) & $F$ &$::=$& $\hole \mid e\,F \mid F\,v$
\end{tabular}
\begin{align*}
 E[\left(\lambda x.e\right)v] & \leadsto E[e\{v/x\}] \\
 E[\letexp{x=v}{e}] &\leadsto  E[e\{v/x\}]\\
 E[\resetp{}{v}] & \leadsto E[v] \\
 E[\resetp{}{F[\shift k.e]}] &\leadsto E[\resetp{}{e\{k:=\lambda y.\resetp{}{F[y]}\}}]
 \qquad y\ \text{is a fresh variable for}\ F
\end{align*}
\caption{Operational Semantics of Source Calculus}\label{fig:source-semantics}
\end{figure}
Evaluation contexts are standard,
and pure evaluation contexts are those evaluation contexts that have
no resets enclosing the hole.
We use the right-to-left evaluation order for the function application
to reflect the semantics of the OCaml compiler.

The first two evaluation rules are the standard beta 
and let rules, where $e\{v/x\}$ denotes capture-avoiding substitution.
The next two rules are those for control operators:
if the body of a reset expression is a value, 
the occurrence of reset is discarded.
If the next redex is a shift expression, we capture
the continuation up to the nearest reset ($\lambda y.\resetp{}{F[y]}$),
and substitute it for $k$ in the body.
$\{k:=\lambda y.\resetp{}{F[y]}\}$ denotes capture-avoiding substitution
for continuation variables where we define
$\throw{k}{e}\{k:=\lambda y.\resetp{}{F[y]}\}$ as
$\resetp{}{(\lambda y.\resetp{}{F[y]})\,(e\{k:=\lambda y.\resetp{}{F[y]}\})}$,
namely $\throw{k}{e}$ is the same as $\resetp{}{k\,e}$ in the original
formulation~\cite{Danvy1990}. The other cases of the substitution are
the same as the standard definition.

Types, type schemes and type environments are defined as follows:
 \begin{align*}
 \tau, \sigma, \alpha, \beta ::=& \  t  \mid b \mid \sigma \rightarrow \tau
 \mid (\sigma/\alpha \rightarrow \tau/\beta) \\
 A ::=& \ \tau \mid \forall t.A\\
 \Gamma ::=&\ \emptyset \mid \Gamma, x:A \mid \Gamma, k:\sigma\to\tau
 \end{align*}
Types are either type variables ($t$), base types $(b)$, pure function
(continuation) type ($\sigma \rightarrow \tau$), or
effectful function types $(\sigma/\alpha \rightarrow \tau/\beta)$,
which represent function types $\sigma \to \tau$ 
where the answer type changes from $\alpha$ to $\beta$. 
Type scheme $A$ represents polymorphic types as usual.
Type environment $\Gamma$ is a finite sequence of variable-type pairs,
which possibly contains continuation variables $k$, that has
a pure function type $\sigma \rightarrow \tau$.

Figure~\ref{fig:shiftreset_with_atm} defines the type system of $\lamATM$.
Type judgments are either $\Gamma \vdash_p e: \tau$ (pure judgments)
or $\Gamma \vdash e: \tau; \alpha, \beta$ (effectful judgments),
the latter of which means that evaluating $e$ with the answer type $\alpha$
yields a value of type $\tau$ with the answer type being modified to
$\beta$. 
The typing rules are based on 
Danvy and Filinski's~\cite{Danvy1989} except that
we have let-polymorphism and clear distinction of pure judgments
from impure judgments following Asai and Kameyama~\cite{Asai2007}.

\begin{figure}[t]
 \begin{minipage}{0.3\hsize}
 \begin{prooftree}
 \AxiomC{$x:A \in \Gamma, \tau < A$}
 \RightLabel{var}
 \UnaryInfC{$\Gamma\vdash_p \  x : \tau$}
 \end{prooftree}
 \end{minipage}
 \begin{minipage}{0.3\hsize}
 \begin{prooftree}
 \AxiomC{($c$ is a constant of type $b$)}
 \RightLabel{const}
 \UnaryInfC{$\Gamma\vdash_p c:b$}
 \end{prooftree} 
 \end{minipage}
 \begin{minipage}{0.3\hsize}
 \begin{prooftree}
  \AxiomC{$\Gamma\vdash_p e:\tau$}
  \RightLabel{exp}
  \UnaryInfC{$\Gamma\vdash e:\tau;~\alpha,~\alpha$}
 \end{prooftree}
 \end{minipage}
 \begin{minipage}{0.4\hsize}
 \begin{prooftree}
 \AxiomC{$\Gamma, x:\sigma \vdash e:\tau;~\beta,~\gamma$}
 \RightLabel{fun}
 \UnaryInfC{$\Gamma \vdash_p \lambda x.e :(\sigma/\beta \rightarrow \tau/\gamma)$}
 \end{prooftree}
 \end{minipage} 
 \begin{minipage}{0.6\hsize}
  \begin{prooftree}
 \AxiomC{$\Gamma \vdash e_1 : (\sigma/\alpha \rightarrow \tau/\beta);~\beta,~\gamma$}
 \AxiomC{$\Gamma \vdash e_2 : \sigma;~\gamma,~\delta$}
 \RightLabel{app}
 \BinaryInfC{$\Gamma \vdash e_1 \, e_2 : \tau ;~\alpha,~\delta$}
 \end{prooftree}
 \end{minipage} 
 \begin{minipage}{0.28\hsize}
 \begin{prooftree}
 \AxiomC{$\Gamma \vdash e:\sigma;~\sigma,~\tau$}
 \RightLabel{reset}
 \UnaryInfC{$\Gamma \vdash_p \left< e\right> : \tau$}
 \end{prooftree}
 \end{minipage}
 \begin{minipage}{0.28\hsize}
 \begin{prooftree}
 \AxiomC{$\Gamma, k:\tau \rightarrow \alpha \vdash_p e : \beta$}
 \RightLabel{shift}
 \UnaryInfC{$\Gamma \vdash \shift k. e: \tau ;~\alpha,~\beta$}
 \end{prooftree}
 \end{minipage} 
 \begin{minipage}{0.44\hsize}
 \begin{prooftree}
 \AxiomC{$\Gamma, k : \sigma \rightarrow \tau \vdash_p e : \sigma$}
 \RightLabel{throw}
 \UnaryInfC{$\Gamma,k : \sigma \rightarrow \tau \vdash_p \throw{k}{e} : \tau$}
 \end{prooftree}
 \end{minipage}
 \begin{minipage}{0.5\hsize}
 \begin{prooftree}
 \AxiomC{$\Gamma \vdash_p v : \sigma$}
 \AxiomC{$\Gamma, x:\mathrm{Gen}\left(\sigma; \Gamma\right) \vdash e :
 \tau;~\alpha,~\beta$}
 \RightLabel{let}
 \BinaryInfC{$\Gamma\vdash \letexp{x=v}{e} : \tau;~\alpha,~\beta$}
 \end{prooftree}
 \end{minipage}
 \caption{Typing Rules of the Source Calculus}\label{fig:shiftreset_with_atm}
\end{figure}

In the var rule, $\tau < A$ means that the type $\tau$ is 
an instance of type scheme $A$, and 
the type $\mathrm{Gen}\left(\sigma; \Gamma\right)$
denotes $\forall t_1,\cdots \forall t_n.\sigma$ where
$t_1,\cdots,t_n$ are the type variables that appear in $\sigma$ but
not appear in $\Gamma$ freely.

The delimited continuations captured by shift expressions are 
pure functions (they are polymorphic in answer types), 
and we use the pure function space $\tau \to \alpha$ for this purpose.  
On the contrary, the functions introduced by lambda are, in general, effectful.
Accordingly, we have two rules for applications.
Note that the body of a shift expression is restricted to a pure expression
in order to simplify the definition of our translation.
This restriction is inessential; 
in the standard formulation (where the body of shift is an effectful
expression), the term 
$\shift x. e$ is typable if and only if
$\shift x. \resetp{}{e}$ is typable,
and their operational behaviors are the same.
The exp rule turns pure terms into effectful terms where
we have chosen an implicit coercion from a pure term to an effectful one.

The type system of the source calculus $\lamATM$ enjoys 
the subject reduction property.
The proof is standard and omitted.

We introduce the target calculus $\mpsr$, which is
a polymorphic calculus with
multi-prompt shift and reset without ATM.
The calculus is similar, in spirit, to 
Gunter et al.'s calculus with the \texttt{cupto} and \texttt{set}
operators~\cite{Gunter1995}.
Besides disallowing ATM, the target calculus differs from the
source calculus in that the control operators are named,
to allow mixing multiple effects in a single program.
The names for control operators are called \textit{prompts} for historical
reasons, and denoted by $p,q,\cdots$.
In our formulation, prompts are first-class values and can be bound
to ordinary variables $x$.  Prompts are given as prompt-constants,
or can be generated dynamically by the $\newp$ primitive. For instance,
evaluating $\newp x.\resetp{x}{1+\shift_{x} k.e}$ first
creates a fresh prompt $p$ and substitutes it for $x$, then it evaluates
$\resetp{p}{1+\shift_{p} k.e}$.
This choice of the formulation closely follows Kiselyov's DelimCC library
for multi-prompt shift and reset.

Types and typing environments are defined as follows:
  \begin{align*}
   \tau, \sigma ::=& \ t \mid b \mid \sigma \rightarrow \tau \mid
   \prompt{\tau}\\
   A ::=& \ \tau \mid \forall t.A\\
   \Gamma ::=& \ \emptyset \mid \Gamma, x:A
  \end{align*}
where $\prompt{\tau}$ is the type for the prompts with the answer type $\tau$.
The syntax of values and terms are defined as follows:
  \begin{align*}
   v ::=& \ x \mid c \mid \lambda x.e \mid p\\
   e ::=& \ v \mid e_1e_2 \mid \shift_{v} x.e \mid \resetp{v}{e} \mid \newp x.e
   \mid \letexp{x = v}{e} \mid \Omega
  \end{align*}
where $p$ is a prompt-constant.  The control operators now receive
not only prompt-constants, but values which will reduce to prompts.
Other values are rejected by the type system.
The term $\newp x.e$ creates a fresh prompt and binds $x$ to it.
The term $\Omega$ denotes a non-terminating computation of arbitrary types. 
It may be defined in terms of shift, 
but for the sake of clarity, we added it as a primitive.

Evaluation contexts and evaluation rules are given as follows:
\begin{align*}
 E & ::= \hole \mid Ee \mid vE \mid \resetp{p}{E}  \\
 E[(\lambda x.e)v] &\leadsto E[e\{v/x\}] \\
 E[\letexp{x = v}{e}] &\leadsto E[e\{v/x\}] \\
 E[\newp x.e] &\leadsto E[e\{p/x\}] \qquad p\ \text{is a fresh prompt-constant} \\
 E[\resetp{p}{v}] &\leadsto E[v] \\
 E[\resetp{p}{E_p[\shift_p x.e]}] &\leadsto
 E[\resetp{p}{e\{\lambda y.\resetp{p}{E_p[y]}/x\}}] \\
 E[\Omega] &\leadsto E[\Omega]
\end{align*}
Note that we use $E_p$ in the second last rule, which is
an evaluation context that does not have a reset with
the prompt $p$ around the hole, and thus implies that
we capture the continuation up to the nearest reset with the prompt $p$.

\begin{figure}[t]
\begin{minipage}{0.3\hsize}
 \begin{prooftree}
 \AxiomC{$x:A \in \Gamma, \tau < A$}
 \RightLabel{var}
 \UnaryInfC{$\Gamma\vdash \  x : \tau$}
 \end{prooftree}
\end{minipage}
\begin{minipage}{0.3\hsize}
\begin{prooftree}
 \AxiomC{($c$ is a constant of $b$)}
 \RightLabel{const}
 \UnaryInfC{$\Gamma\vdash c:b$}
\end{prooftree} 
\end{minipage}
\begin{minipage}{0.3\hsize}
 \begin{prooftree}
  \AxiomC{\ \phantom{a} \vspace{1em}}
  \RightLabel{omega}
  \UnaryInfC{$\Gamma \vdash \Omega:\tau$}
 \end{prooftree}
\end{minipage}

\begin{minipage}{0.5\hsize}
 \begin{prooftree}
 \AxiomC{$\Gamma, x:\sigma \vdash e:\tau$}
 \RightLabel{fun}
 \UnaryInfC{$\Gamma \vdash \lambda x.e :\sigma \rightarrow \tau$}
 \end{prooftree}
\end{minipage}
\begin{minipage}{0.5\hsize}
  \begin{prooftree}
 \AxiomC{$\Gamma \vdash e_1 : \sigma \rightarrow \tau$}
 \AxiomC{$\Gamma \vdash e_2 : \sigma$}
 \RightLabel{app}
 \BinaryInfC{$\Gamma \vdash e_1 \, e_2 : \tau$}
 \end{prooftree}
\end{minipage} 

\begin{minipage}{0.5\hsize} 
 \begin{prooftree}
 \AxiomC{$\Gamma \vdash v : \prompt{\tau}$}
 \AxiomC{$\Gamma \vdash e:\tau$}
 \RightLabel{reset}
 \BinaryInfC{$\Gamma \vdash \resetp{v}{e}: \tau$}
 \end{prooftree}
\end{minipage}
\begin{minipage}{0.5\hsize}
 \begin{prooftree}
 \AxiomC{$\Gamma \vdash v: \prompt{\sigma}$} 
 \AxiomC{$\Gamma, x:\tau \rightarrow \sigma \vdash e : \sigma$}
 \RightLabel{shift}
 \BinaryInfC{$\Gamma \vdash \shift_{v} \ x. e: \tau$}
 \end{prooftree}
\end{minipage} 

\begin{minipage}{0.5\hsize}
 \begin{prooftree}
 \AxiomC{$\Gamma \vdash v : \sigma$}
 \AxiomC{$\Gamma, x:\mathrm{Gen}\left(\sigma; \Gamma\right) \vdash e:\tau$}
 \RightLabel{let}
 \BinaryInfC{$\Gamma\vdash \letexp{x = v}{e} : \tau$}
 \end{prooftree}
\end{minipage}
\begin{minipage}{0.5\hsize}
 \begin{prooftree}
  \AxiomC{$\Gamma, x: \prompt{\sigma} \vdash e:\tau $}
  \RightLabel{prompt}
  \UnaryInfC{$\Gamma \vdash \newp x.e:\tau$}
 \end{prooftree}
\end{minipage}

  \caption{Typing Rules of the Target Calculus}\label{fig:mp-sr}
 \end{figure}

Finally we give typing rules for the target calculus in 
Figure~\ref{fig:mp-sr}.  
The type system of the target calculus is mostly standard except 
for the use of prompts.  In the shift rule, the prompt expression
$v$ must be of type $\prompt{\sigma}$ where $\sigma$ is the type
of the body of the shift expression.
A similar restriction is applied to the reset rule.
In the prompt rule, we can create an arbitrary prompt and bind
a variable $x$ to it.

The type system enjoys the subject reduction property modulo the 
set of dynamically created prompts which have infinite extents.

%% file: translation.tex

\section{The PPS Translation}\label{sec:trans}

In this section, we give a Prompt-Passing Style (PPS) translation, 
the syntax-directed translation from 
$\lamATM$ to $\mpsr$, which translates away the feature of
answer-type modification. 

The translation borrows the idea of Kiselyov's implementation
of typed printf in terms of multi-prompt shift and reset, but 
this paper gives a translation for the whole calculus and 
also proves the type preservation property. 
Later, we will show a tagless-final
implementation based on our translation which is another evidence that
our translation actually preserves typing.

Figure~\ref{fig:trans_type} defines the translation rules
for types, type schemes, type environments and triples.

\begin{figure}
 \begin{align*}
 \jump{\tau; \alpha, \beta} &= \prompt{\jump{\beta}} \rightarrow
 \prompt{\jump{\alpha}} \rightarrow \jump{\tau} \\
 \jump{b} &= b \\
 \jump{t} &= t \\
 \jump{\sigma \rightarrow \tau} &= \jump{\sigma} \rightarrow
 \jump{\tau} \\
 \jump{\sigma/\alpha \rightarrow \tau/\beta} &= \jump{\sigma} \rightarrow
 \jump{\tau; \alpha, \beta} \\
  \jump{\forall t.A} &= \forall t.\jump{A} \\
  \jump{\emptyset} &= \emptyset \\
 \jump{\Gamma, x:A} &= \jump{\Gamma}, x:\jump{A}\\
  \jump{\Gamma, k:\sigma \to \tau} &= \jump{\Gamma}, k:\jump{\sigma} \to \jump{\tau} 
\end{align*}
\caption{Translation for Triples, Types, Type Schemes and Type Environments}\label{fig:trans_type}
\end{figure}

As we have explained in earlier sections, we emulate ATM from
the type $\alpha$ to the type $\beta$ in terms of
two prompts whose answer types are $\prompt{\alpha}$ and $\prompt{\beta}$.
Hence the triple $\tau; \alpha, \beta$ in the typing judgment
is translated to the type 
$\prompt{\jump{\beta}} \to \prompt{\jump{\alpha}} \to \jump{\tau}$.

Types are translated in a natural way
except the type for effectful functions
$\sigma/\alpha \rightarrow \tau/\beta$; it is
translated to a function type whose codomain
is the translation of the triple $\jump{\tau; \alpha, \beta}$.
Type schemes and type environments are translated naturally.

\begin{figure}[t]
 \begin{align*}
 \puretr{x} &= x \\
 \puretr{c} &= c \\
 \puretr{\lambda x.e} &= \lambda x.\jump{e} \\
 \puretr{\throw{k}{e}} &= k\,\puretr{e} \\
 \puretr{\resetp{}{e}} &= \newp pq. \resetp{p}{(\lambda y.\shift_q \_.y)(\jump{e}p q)} \\
 \jump{e_1e_2} &= \lambda pq. \newp rs. (\jump{e_1}r s)
 (\jump{e_2}p r) s q\\
 \jump{\letexp{x=v}{e_2}}&= \lambda pq. \letexp{x = \puretr{v}}{\jump{e_2}pq}\\
 \jump{\shift k.e} &= \lambda p q. \shift_p
 k'. ((\lambda k.\puretr{e})(\lambda y.\resetp{q}{(\lambda \_.\Omega)(k'
  y)}) \\
 \jump{e} &= \lambda p q. \shift_p k.\resetp{q}{k \puretr{e}} 
 \qquad \text{if}~ e = x,~c,~\lambda x.e',~
\resetp{}{e'}~\text{or}~ \throw{k}{e'}
\end{align*}
 \caption{Translation for Typed Terms}\label{fig:trans_term}
\end{figure}

Figure~\ref{fig:trans_term} defines the translation 
for typed terms in $\lamATM$, which consists of
the translation $\puretr{e}$ for a pure term $e$,
and the translation $\jump{e}$ for an effectful term $e$.
These translations are defined for typed terms only, since
the translation for $\throw{k}{e}$ (and $\shift k.e$) contains
$\puretr{e}$, which is defined only for a pure term $e$.

The first translation $\puretr{e}$ does little for most constructs
but reset terms.
A reset term $\resetp{}{e}$
is translated to a term which creates two new prompts $p$ and $q$,
and inserts a combination of reset-p and shift-q as we explained earlier.
Then it supplies $p$ and $q$ to its immediate subterm $\jump{e}$.

The second translation $\jump{e}$ does a lot;
For application $e_1\,e_2$, it receives two prompts $p$ and $q$,
but also generates two new prompts $r$ and $s$ and distributes
these prompts to its subterms etc.
For a let-term $\letexp{x=v}{e}$, it passes two prompts.

The translation of a shift term $\shift k.e$ is a way more complicated
than others;
it receives two prompts $p$ and $q$ and invokes shift to capture the
delimited continuation and binds $k'$ to it.
The captured delimited continuation is slightly different from the
one which would have been obtained by the shift-operator in the source term, 
since we have
inserted a combination of reset-p and shift-q at the position of reset.
As we have explained in Section \ref{sec:simulation},
this (bad) effect is resolved by detoxifying the continuation, which is
realized by the involved term found in the translation above.

The last clause for $\jump{e}$ applies only when 
the type of the term $e$ is derived by applying the exp rule as its
last rule.   
In this case, the translation generates a term 
which receives two prompts $p$ and $q$, detoxify the effect of the delimited
continuation (mentioned above) by the combination of shift-p and reset-q.

We can show that our translation preserves typing.

\begin{theo}[Type preservation]
If $\Gamma \vdash e: \tau; \alpha, \beta$ is derivable in the source
calculus $\lamATM$,
then $\jump{\Gamma} \vdash \jump{e} :\jump{\tau; \alpha, \beta}$ is
derivable in the target calculus $\mpsr$. 

Similarly, 
if $\Gamma \vdash_p e:\tau$ is derivable in $\lamATM$,
so is $\jump{\Gamma} \vdash \puretr{e} :\jump{\tau}$ in $\mpsr$.
\end{theo}

Proof. We will prove the two statements by simultaneous induction 
on the derivations.
Here we only show a few interesting cases.

(Case $e=\resetp{}{e_1}$)
We have a derivation for:
\begin{prooftree}
 \AxiomC{$\Gamma \vdash e_1:\sigma;\sigma, \tau$}
 \UnaryInfC{$\Gamma \vdash_p \resetp{}{e_1} : \tau$}
\end{prooftree}
By induction hypothesis,
we can derive 
$\jump{\Gamma} \vdash \jump{e_1}:\jump{\sigma; \sigma, \tau}$.
Let
$\Gamma'=\jump{\Gamma}, p:\prompt{\jump{\tau}}, q:\prompt{\jump{\sigma}}$,
$\Gamma'' = \Gamma', y:\jump{\sigma}$, and $\Gamma''' = \Gamma'', x:\jump{\tau}\to\jump{\sigma}$.
We have the following derivation:
{\footnotesize
\begin{prooftree}
   \AxiomC{$\Gamma' \vdash p :\prompt{\jump{\tau}}$}

       \AxiomC{$\Gamma''' \vdash q:\prompt{\jump{\sigma}}$}
       \AxiomC{$\Gamma''' \vdash y:\jump{\sigma}$}
     \BinaryInfC{$\Gamma'' \vdash \shift_q x .y : \jump{\tau}$}
   \UnaryInfC{$\Gamma' \vdash \lambda y.\shift_q x .y : \jump{\sigma} \rightarrow \jump{\tau}$}

     \AxiomC{$\Gamma' \vdash q:\prompt{\jump{\sigma}}$}
       \AxiomC{$\Gamma' \vdash p:\prompt{\jump{\tau}}$}
       \AxiomC{$\Gamma' \vdash \jump{e_1}: \jump{\sigma; \sigma,\tau}$}
     \BinaryInfC{$\Gamma' \vdash \jump{e_1}p: \prompt{\jump{\sigma}} \rightarrow \jump{\sigma}$}
   \BinaryInfC{$\Gamma' \vdash \jump{e_1}pq : \jump{\sigma}$}

  \BinaryInfC{$\Gamma' \vdash \left(\lambda y. \shift_q x .y\right)\left(\jump{e_1}pq\right) : \jump{\tau}$}
\BinaryInfC{$\Gamma' \vdash \resetp{p}{\left(\lambda y. \shift_q x .y\right)\left(\jump{e_1}pq\right)}:\jump{\tau}$}

\doubleLine
\UnaryInfC{$\jump{\Gamma}\vdash \newp{}p.\newp{}q.\resetp{p}{\left(\lambda v. \shift_q x .v\right)\left(\jump{e_1}pq\right)}:\jump{\tau}$}
\end{prooftree}
}
which derives 
$\jump{\Gamma} \vdash \puretr{\resetp{}{e_1}}:\jump{\tau}$.

(Case $e=\shift x.e_1$)
We have a deviation for
\begin{prooftree}
 \AxiomC{$\Gamma, k:\tau \rightarrow\alpha \vdash_p e_1:\beta$}
 \UnaryInfC{$\Gamma \vdash \shift k.e_1:\tau; \alpha, \beta$}
\end{prooftree}
By induction hypothesis
$\jump{\Gamma, k:\tau\rightarrow\alpha}\vdash \puretr{e_1}:\jump{\beta}$
is derivable.
Let
$\Gamma' = \jump{\Gamma}, p:\prompt{\jump{\beta}}, q:\prompt{\jump{\alpha}}$,
$\Gamma'' = \Gamma', k':\jump{\tau}\rightarrow\jump{\beta}$, and
$\Gamma''' = \Gamma'', y:\jump{\tau}$,
then we have:
\begin{prooftree}
\AxiomC{$\Gamma'' \vdash p:\prompt{\jump{\beta}}$}
 \AxiomC{$\Gamma'', k:\jump{\tau}\rightarrow\jump{\alpha} \vdash \puretr{e_1}: \jump{\beta}$}
 \UnaryInfC{$\Gamma'' \vdash \lambda k.\puretr{e_1}: \left(\jump{\tau}\rightarrow\jump{\alpha}\right)\rightarrow \jump{\beta}$}
 \AxiomC{$\Gamma''' \vdash \resetp{q}{\left(\lambda \_ .\Omega\right)\left(k'y\right)}: \jump{\alpha}$} 
 \UnaryInfC{$\Gamma'' \vdash \lambda y. \resetp{q}{\left(\lambda \_ .\Omega\right)\left(k'y\right)}:
             \jump{\tau}\rightarrow\jump{\alpha}$}
 \BinaryInfC{$\Gamma'' \vdash \left(\lambda k.\puretr{e_1} \right) \left(\lambda y. \resetp{q}{\left(\lambda \_ .\Omega\right)
              \left(k'y\right)}\right): \jump{\beta}$}
 \BinaryInfC{$\Gamma' \vdash \shift_p k'.\left(\lambda k.\puretr{e_1} \right) \left(\lambda y. \resetp{q}{\left(\lambda \_ .\Omega\right)\left(k'y\right)}\right):\jump{\tau}$}
 \doubleLine
 \UnaryInfC{$\jump{\Gamma} \vdash
             \lambda p.\lambda q.\shift_p k'.\left(\lambda k.\puretr{e_1} \right) \left(\lambda y. \resetp{q}{\left(\lambda \_ .\Omega\right)\left(k'y\right)}\right):
             \prompt{\jump{\beta}}\rightarrow\prompt{\jump{\alpha}}\rightarrow\jump{\tau}$}
\end{prooftree}
which derives $\jump{\Gamma} \vdash \jump{\shift k.e_1}:
\jump{\tau; \alpha, \beta}$.  ~\hfill$\Box$

Hence our translation preserves typing.
We conjecture that our translation also preserve operational semantics but
its proof is left for future work.


%% file: opt.tex

\section{Optimization} \label{sec:opt}

The naive PPS translation in Section \ref{sec:trans} works in theory, but is suboptimal for
practical use, as it introduces too many prompts.  To solve this problem,
we will introduce an optimized PPS translation in this section.

For the purpose of comparison,
we consider the term $\resetp{}{e_n}$ for a natural number $n$,
where $e_n$ is $1+(2+\cdots+(n+\shift k.\lambda x.\throw{k}{x})\cdots)$.%
\footnote{The term $\shift k.\lambda x.\throw{k}{x}$ is written as
$\shift k.k$ in the standard formulation. }
Then we can derive $\vdash e_n : \Int;~\Int,~(\Int/\alpha\to\Int/\alpha)$ and
$\vdash_p \resetp{}{e_n} : (\Int/\alpha\to\Int/\alpha)$ 
for some type $\alpha$, in the type system of $\lamATM$ 
augmented by the following type rule for addition:
 \begin{prooftree}
 \AxiomC{$\Gamma \vdash e_1 : \Int ;~\alpha,~\gamma$}
 \AxiomC{$\Gamma \vdash e_2 : \Int ;~\gamma,~\beta$}
 \BinaryInfC{$\Gamma \vdash e_1 + e_2 : \Int ;~\alpha,~\beta$}
 \end{prooftree}
The type rule for addition in the target calculus is standard and omitted.

We define the (naive) PPS translation for addition
by $\jump{e_1+e_2} \equiv \lambda p. \lambda q.
\newp{r}.~(\jump{e_1}rq) + (\jump{e_2}pr)$.
It is easy to see that the naive PPS translation in the previous section
translates $\resetp{}{e_n}$ to a rather big term which 
dynamically generates $n+2$ prompts ($2$ for reset, and $1$ for 
each addition) and passes them around. This is not ideal and needs improvement.

\textbf{Eliminating Unnecessary Prompt Passing}

We first eliminate unnecessary dynamic passing of prompts.
Since the residual terms of the naive translation 
often contain $\jump{e}pq$ as subterms where
$\jump{e}$ takes the form $\lambda p.\lambda q.\cdots$, 
they contain many beta redexes (administrative redexes)
that can be eliminated at the translation time, 
by adjusting 
the one-pass CPS translation by Danvy and Filinski~\cite{Danvy1992}
to our PPS translation.

Figure~\ref{fig:one-pass} gives our one-pass PPS translation 
where function applications are made explicit (by the infix symbol @),
and the overline (e.g. $\overline{@}$)
means static constructs which are reduced at the 
translation time, while the underline (e.g. $\underline{@}$)
means dynamic constructs which remain in the residual terms.

\begin{figure}
 \begin{align*}
 \puretrone{x} &= x \\
 \puretrone{c} &= c \\
 \puretrone{\lambda x.e} &= \dlam{x}{\dlamtwo{p}{q}{\sapptwo{\jumpone{e}}{p}{q}}} \\
 \puretrone{\throw{k}{e}} &= \dapp{k}{\puretrone{e_2}} \\
 \puretrone{\resetp{}{e}} &= \dnewptwo{p}{q}{\resetpu{p}{
       \dapp{(\dlam{y}{\dshift{q}{\_}{y}})}{(\sapptwo{\jumpone{e}}{p}{q})}}} \\
 \jumpone{e_1+e_2} &= \slamtwo{p}{q}{\dnewp{r}{
           (\sapptwo{\jumpone{e_1}}{r}{q}) +
           (\sapptwo{\jumpone{e_2}}{p}{r})
        }} \\
 \jumpone{e_1\,e_2} &= \slamtwo{p}{q}{\dnewptwo{r}{s}{
\dapptwo{
           \dapp{(\sapptwo{\jumpone{e_1}}{r}{s})}
                {(\sapptwo{\jumpone{e_2}}{p}{r})}
        }{s}{q}}} \\
 \jumpone{\letexp{x=v}{e_2}}&= \slamtwo{p}{q}{
         \letexpu{x = \puretrone{v}}{\sapptwo{\jumpone{e_2}}{p}{q}}
       }\\
 \jumpone{\shift k.e} &= 
        \slamtwo{p}{q}{\dshift{p}{k'}{
           \dapp{ (\dlam{k}{\puretrone{e}}) }
                {\left(\dlam{y}{\resetpu{q}{
                    \dapp{(\dlam{\_}{\Omega})}
                         {(\dapp{k'}{y})}}}
                 \right)
                }}}  \\
 \jumpone{e} &= \slamtwo{p}{q}{\dshift{p}{k}{
                 \resetpu{q}{\dapp{k}{\puretrone{e}}}
              }} 
 \qquad \text{if}~ e = x,~c,~\lambda x.e',~
\resetp{}{e'}~\text{or}~ \throw{k}{e'}
 \end{align*}
\caption{One-Pass PPS Translation} \label{fig:one-pass}
\end{figure}

The one-pass PPS translation eliminates unnecessary prompt passing;
Applying one-pass PPS translation to a term in $\lamATM$,
and reducing all static beta-redexes $\sapp{(\slam{p}{e_1})}{q}$
to $e_1\{q/p\}$, one obtains a term without static constructs
(constructs with overlines).
The residual terms of one-pass PPS translation pass prompts only for
function applications; in Figure~\ref{fig:one-pass},
dynamic application for prompts $\dapptwo{e}{p}{q}$ appears in the
term $\jumpone{e_1\,e_2}$ only.


The one-pass PPS translation gives a much better result than the naive one,
as it eliminates unnecessary prompt passing, but
it still generates as many prompts as the naive one.

\textbf{Eliminating Unnecessary Prompt Generation}

We eliminate unnecessary prompt generation. Our idea is simple;
if the term being translated is pure, we do not have to generate prompts.
In the translation of $\jump{e_1+e_2} =
\lambda p.\lambda q.\newp{r}.~(\jump{e_1}rq) + (\jump{e_2}pr)$,
the new prompt $r$ is used to bridge 
$\jump{e_1}$ and $\jump{e_2}$, and if one of $e_1$ and $e_2$ is
pure, we can reuse $p$ and $q$ to simulate ATM.
For instance, if $e_1$ is pure and $e_2$ is effectful, 
we can define $\jump{e_1+e_2} =
\lambda p.\lambda q.\puretr{e_1} + (\jump{e_2}pq)$.

To maximize the benefit of this optimization, we 
extend the notion of a pure term $e_p$ to a \textit{quasi-pure} term 
(or a q-pure term) $e_q$ by:
\begin{align*}
e_p & ::= x \mid c \mid \lambda x.e \mid \resetp{}{e} \mid \throw{k}{e} \\
e_q & ::= e_p \mid \letexp{x=v}{e_q}
\end{align*}
Namely, we allow nested let constructs appearing around pure terms.
For instance, $\letexp{x=3}{\letexp{y=5}{7}}$ is not pure, but
is q-pure. 

Figure~\ref{fig:opt} defines the optimized PPS translation;
$\puretrtwo{e}$ for a q-pure term $e$, and 
$\jumptwo{e}$ for a non q-pure term $e$.
We omit the cases whose translation is the same as 
those for one-pass PPS translation.
\begin{figure}
 \begin{align*}
 &\puretrtwo{\lambda x.e_1} = 
    \left\{
      \begin{array}{ll}
        \dlam{x}{\dlamtwo{p}{q}{\dshift{p}{k}{
                                  \resetpu{q}{\dapp{k}{\puretrtwo{e_1}}}
                              }}}
             &  \quad \text{if}~~ e_1 ~\text{is q-pure} \\[5pt]
        \dlam{x}{\dlamtwo{p}{q}{\sapptwo{\jumptwo{e}}{p}{q}}}
             &  \quad \text{otherwise} 
      \end{array}
    \right. \\
 &\puretrtwo{\resetp{}{e_1}} = 
    \left\{
      \begin{array}{ll}
        \puretrtwo{e_1} & \quad \text{if}~~ e_1 ~\text{is q-pure} \\
        \dnewptwo{p}{q}{\resetpu{p}{
           \dapp{(\dlam{y}{\dshift{q}{\_}{y}})} 
                {(\sapptwo{\jumptwo{e_1}}{p}{q})}
           }}
           & \quad \text{otherwise}  
      \end{array}  
    \right. \\
 &\jumptwo{e_1\,e_2} = 
    \left\{
      \begin{array}{ll}
 \slamtwo{p}{q}{\dapptwo{\dapp{\puretrtwo{e_1}}{\puretrtwo{e_2}}}{p}{q}}
            & \text{if}~~e_1 ~\text{and}~e_2~\text{are q-pure} \\
 \slamtwo{p}{q}{\dnewp{r}{
         \dapptwo{\dapp{\puretrtwo{e_1}}{(\sapptwo{\jumptwo{e_2}}{p}{r})}
                 }{r}{q}
         }}
            & \text{if}~~e_1 ~\text{is q-pure}
                  ~~\text{and}~~e_2 ~\text{is not} \\
 \slamtwo{p}{q}{\dnewp{r}{
         \dapptwo{\dapp{ (\sapptwo{\jumptwo{e_1}}{p}{r}) }{\puretrtwo{e_2}}}
                 {r}{q}
       }}
            & \text{if}~~e_2 ~\text{is q-pure}
                  ~~\text{and}~~e_1 ~\text{is not} \\
 \slamtwo{p}{q}{\dnewptwo{r}{s}{
       \dapptwo{ \dapp{ (\sapptwo{\jumptwo{e_1}}{r}{s}) }
                      { (\sapptwo{\jumptwo{e_2}}{p}{r}) }
               }{s}{q}
            }}
            & \text{otherwise}
      \end{array}  
    \right. \\
 &\puretrtwo{\letexp{x=v}{e_1}} = 
  \letexpu{x = \puretrtwo{v}}{\puretrtwo{e_1}} 
                 \quad \text{if}~~e_1 ~\text{is q-pure} \\
  &\jumptwo{\letexp{x=v}{e_1}} = 
  \slamtwo{p}{q}{ \letexpu{x = \puretrtwo{v}}{\sapptwo{\jumptwo{e_1}}{p}{q}} }
                \quad \text{if}~~e_1 ~\text{is not q-pure}
 \end{align*}
\caption{Optimized PPS Translation (new cases only)} \label{fig:opt}
\end{figure}
The optimized PPS translation dispatches if each subterm is q-pure or not,
and in the former case, it gives an optimized result where prompt generation
is suppressed.
Translation for addition can be defined similarly, for instance,
$\jumptwo{e_1+e_2} = 
\slamtwo{p}{q}{\puretrtwo{e_1} + (\sapptwo{\jumptwo{e_2}}{p}{q})}$
if $e_1$ is q-pure and $e_2$ is not.

We compute $\jump{\resetp{}{e_n}}$ 
and reduce all static redexes in it, 
to obtain the following term:\footnote{We omitted the underlines in the result.}
\[
\newp{p}.\newp{q}.~
\resetp{p}{
(\lambda y.\shift_q \_.y)
(1+(2+\cdots+(n+ (\shift_p k'.(\lambda k.
\puretrtwo{\lambda x.\throw{k}{x}})@(\lambda y.\resetp{q}{\dots})))\cdots))}
\]
where only two prompts are generated at the beginning of the computation.
The result is quite close to the source term, and in fact,
the source and target terms differ 
only at the control operators.
Although it is possible to further optimize the results,
by applying partial evaluation techniques for shift and reset 
(e.g. \cite{Asai:PE}),
we believe that our translation is practical and efficient.
In the next section, we show an implementation of our translation based
on the optimized PPS translation.

We can prove that the optimized PPS translation preserves types 
where the type system of the target calculus $\mpsr$
has two function types $\sigma \overline{\to} \tau$ (static) and
$\sigma \underline{\to} \tau$ (dynamic).
Namely, we can prove that,
if $\Gamma \vdash e: \tau; \alpha, \beta$ is derivable in $\lamATM$ and
$e$ is not q-pure, so is
$\jumptwo{\Gamma} \vdash \jumptwo{e} :
\prompt{\jumptwo{\beta}} \overline{\to}
\prompt{\jumptwo{\alpha}} \overline{\to} \jumptwo{\tau}$ in $\mpsr$.
Similarly,
if $\Gamma \vdash e: \tau; \alpha, \alpha$ 
is derivable and $e$ is q-pure, or
$\Gamma \vdash_p e: \tau$ is derivable, so is
$\jumptwo{\Gamma} \vdash \puretrtwo{e} : \jumptwo{\tau}$.
The details of this development and proofs are omitted.


%% file: tagless_final.tex

\section{Tagless-final embedding} \label{sec:tagless_final}

We have implemented the calculus $\lamATM$ in 
Figure~\ref{fig:shiftreset_with_atm}
and the naive and optimized PPS translations in
Figures~\ref{fig:trans_type}, \ref{fig:trans_term} and \ref{fig:opt}
for a monomorphic version of $\lamATM$.

Our implementation is based on the tagless-final 
style~\cite{Carette2009,Kiselyov:2010:TTF}, 
which allows one to embed a typed domain-specific language (DSL) 
in a metalanguage.
In this style, the syntax as well as the typing rules of DSL 
are represented by a signature (an interface of modules),
and its semantics is given as an interpretation of this signature.
One of the important merits with this style is that 
type checking (or type inference) of DSL is automatically
done by the type checker (or the type inferencer) of the metalanguage.
Although we have already proved the subject reduction property of $\lamATM$ and
the type preservation property for the PPS translations, 
implementing them in the tagless-final style gives us 
another indication for well typedness. It is particularly useful 
when we extend the source calculus and the translation; type errors are
immediately raised by the type system of the metalanguage.

We have chosen 
OCaml plus the DelimCC library
as the metalanguage,
where DelimCC gives an efficient implementation
for multi-prompt shift and reset~\cite{Kiselyov2012}.
We also give an implementation in MetaOCaml, a multi-stage extension of OCaml, 
to generate (and show) the translated terms, 
rather than immediately executing them.

\begin{figure}[t]
 \begin{lstlisting}[style=ocaml]
module type Symantics = sig
  type 't pure               (* pure expression *)
  type ('t, 'a, 'b) eff        (* effectful expression *)
  type ('s, 't, 'a, 'b) efun    (* effectful function type *)
  type ('s, 't) pfun          (* pure function type *)
  val const : 't -> 't pure
  val lam : ('s pure -> ('t, 'a, 'b) eff) -> ('s, 't, 'a, 'b) efun pure
  val app : (('s, 't, 'a, 'b) efun, 'b, 'c) eff
       -> ('s, 'c, 'd) eff -> ('t, 'a, 'd) eff
  val throw : ('s, 't) pfun pure -> 's pure -> 't pure
  val shift : (('t, 'a) pfun pure -> 'b pure) -> ('t, 'a, 'b) eff
  val reset : ('s, 's, 't) eff -> 't pure
  val exp : 't pure -> ('t, 'a, 'a) eff
  val run : 't pure -> 't
end
\end{lstlisting}
 \caption{Signature of the Embedded Language}
 \label{fig:signature}
\end{figure}

Figure~\ref{fig:signature} shows the signature 
called \lstinline{Symantics} for our source calculus $\lamATM$.
It represents the syntax and the typing rules of $\lamATM$;
the types \lstinline[style=ocaml]|'t pure| and 
\lstinline[style=ocaml]|('t, 'a, 'b) eff|, resp.,
represent the relations $\Gamma \vdash_p e : \tau$ and
$\Gamma \vdash e : \tau;~\alpha,~\beta$, resp.
The type \lstinline[style=ocaml]|('s, 't, 'a, 'b) efun| represents 
the effectful function type $(\sigma/\alpha\to\tau/\beta)$
and \lstinline[style=ocaml]|('s, 't) pfun| the pure function type $\sigma\to\tau$
for continuations.
Since all these types are kept abstract, we can arbitrarily instantiate them
in different implementations.
Each function but \lstinline[style=ocaml]|run|
encodes a typing rule in $\lamATM$.
For instance,
the function \lstinline[style=ocaml]|exp| 
encodes the exp rule in $\lamATM$, and 
does not have a concrete primitive in DSL.
The function \lstinline[style=ocaml]|run| does not correspond to a constructor in
DSL; it converts a DSL value to a value in the metalanguage, and is thus
useful to test interpreters.

As an example, a DSL term $\shift k.\lambda x.\throw{k}{x}$
is represented by the term\\
\lstinline[style=ocaml]|shift (fun k -> lam (fun x -> exp (throw k x)))|,
which encodes a type derivation of the above DSL term using 
higher-order abstract syntax.
Note that, we can represent all and only typable terms in $\lamATM$ 
using this signatures, and the typability of
embedded terms are checked by OCaml; 
all representable terms are typable \textit{as they are constructed}.

In the tagless-final style,
operational semantics of the embedded language
is given as an interpretation of the Symantics signature, 
namely, a module of type Symantics.
For this work, we have given 
two interpretations for each of two PPS translations, and thus obtained
four interpreters.
The two interpretations differ in the target; the first one,
called the R interpreter, 
translates the source term and evaluates the result.
The second one, called the S interpreter\footnote{The S interpreter 
is actually a \textit{compiler}.}
translates the source term and generates the result as a code in
MetaOCaml, which can be executed by the \lstinline[style=ocaml]|run| primitive.

Due to lack of space, we cannot list the source code of these interpreters,
but it should be noted that the two PPS translations (the naive one and
the optimized one) have been successfully implemented in the tagless-final
style.
After extending the source calculus with conditional, recursion and
so on, 
we can write programming examples such as list-append in Section~\ref{sec:intro},
list-prefix and others, 
and running these examples gives correct answers.

Figure~\ref{fig:prog_ex} shows a few results of the optimized translation
with the S interpreter. 
We first define \lstinline[style=ocaml]|append|
and a test program \lstinline[style=ocaml]|res1|.
Then we translate \lstinline[style=ocaml]|res1| and run it, to obtain
the desired list.
We then translate \lstinline[style=ocaml]|append| itself (but not run it),
to obtain the code \lstinline[style=ocaml]|.<let rec g_56 ...>.|
where
the variable \lstinline[style=ocaml]|g_56| corresponds to\footnote{
MetaOCaml renames all bound variables.}
the \lstinline[style=ocaml]|append| function in Section~\ref{sec:intro}.
The result is instructive; 
control operators\footnote{
\texttt{Delimcc.shift} is shift and 
\texttt{Delimcc.push\_prompt} is reset.}
are used only at the point where shift was there in the source term,
in particular, no dynamic prompt generation happens during the 
recursive calls for append.
This result clearly shows the merit of our optimized translation
over the naive translation as well as the definitional CPS translation
~\cite{Danvy1990}.
		
\begin{figure}[h]
\begin{lstlisting}[style=ocaml]
module Example (S: SymPL) = struct
  open S
  let append = fixE (fun f x ->
      ifE (null @@ exp x) (shift (fun k -> k))
	                  (head (exp x) @* app (exp f) (tail @@ exp x)))

  let res1 = run @@
             throw (reset (app (exp append) (exp @@ list [1;2;3])))
		   (list [4;5;6])
end
 
# let _ = let module M = Example(SPL_opt) in M.res1;;
- : int list = [1; 2; 3; 4; 5; 6]
# let _ = let module M = Example(SPL_opt) in M.append;;
- : (int list, int list, int list, (int list, int list) SPL_opt.pfun)
    SPL_opt.efun SPL_opt.pure
= .<
let rec g_56 x_57 p_58 q_59 =
  if x_57 = []
  then
    Delimcc.shift p_58
      (fun k'_62  ->
         (fun x_64  -> x_64)
           (fun y_63  ->
              Delimcc.push_prompt q_59
                (fun ()  ->
                   (fun _  -> Pervasives.failwith "Omega") @@ (k'_62 y_63))))
  else
    (let v2_60 = g_56 (List.tl x_57) p_58 q_59 in
     let v1_61 = List.hd x_57 in v1_61 :: v2_60) in
g_56>. 
\end{lstlisting}
 \caption{Programming Examples}\label{fig:prog_ex}
\end{figure}

This implementation also
provides a good evidence that our translation is type preserving.
Thanks to the tagless-final style, our implementation is extensible,
and in fact, it was easy to add primitives such as the fixpoint operator
to our source language in a type-safe way.

%% file: example.tex
\section{The List-prefix Example}

The append function in Section~\ref{sec:atm} is not the only 
interesting example which uses shift and reset.
Danvy~\cite{Danvy:ILFL88} represented various programming examples
using these control operators.
In this section we take the list-prefix example from his work, and show that it 
is correctly translated and implemented in our system. 

The list-prefix function \verb|prefix| takes a list as its input,
and returns the list of all prefixes of the input.
For example, \verb|prefix [1;2;3]|
yields the list \verb|[[1]; [1;2]; [1;2;3]]|.
Using the control operators shift and reset, \verb|prefix| can be 
implemented as follows:

\begin{lstlisting}[style=ocaml]
let prefix l =
  let rec aux = function
    | [] -> shift (fun k -> [])
    | x :: xs -> x :: shift (fun k -> k [] :: reset (k (aux xs)))
in reset (aux l)
\end{lstlisting}

This implementation is interesting for two reasons; first,
it traverses the input only once. Second, it creates
no intermediate lists (explicitly). One can easily imagine that a
straightforward implementation of the function would not
satisfy these two properties.

Let us infer the type of the inner function \verb|aux|.
The pattern match has two cases, and on the second branch,
the answer type before the execution of this shift is
\lstinline[style=ocaml]|'a list|, since 
the function returns \lstinline[style=ocaml]|x :: shift (fun k -> ...)|,
which has a list type.
Then, the continuation captured by this shift has the type
\lstinline[style=ocaml]|'a list -> 'a list|, which implies that
\lstinline[style=ocaml]|k []| has the type
\lstinline[style=ocaml]|'a list|.
Finally,
the answer type after the execution of this shift 
is \lstinline[style=ocaml]|'a list list|, since it returns
\lstinline[style=ocaml]|(k []) :: ...|.
We have just observed that the answer type changed from 
\lstinline[style=ocaml]|'a list| to 
\lstinline[style=ocaml]|'a list list| during the execution 
of the second branch of \verb|aux|.

The functions \verb|aux| and \verb|prefix| are typable in
$\lamATM$ with a few extensions such as lists and recursion.
In fact, our tagless-final implementation can cope with
these extensions and we only have to rewrite pattern matching by
conditionals.
Our implementation of the function \verb|prefix| is shown below:

\begin{lstlisting}[style=ocaml]
module Ex2 (S: SymPL) = struct
 open S
 let prefix = fixE (fun f x ->
   ifE (null @@ exp x)
     (shift (fun k -> list []))
     (head (exp x) @* shift (fun k ->
       reset @@ (exp (throw k (list [])))
       @* (exp (reset @@ app (exp @@ lam (fun x -> exp @@ throw k x))
			     (app (exp f) (tail @@ exp x)))))))
 let res = run @@ reset @@ app (exp prefix) (exp @@ list [1;2;3])
end
\end{lstlisting}

The module \verb|Ex2| contains the definition of \verb|prefix| and
its example use \verb|res|.  The type of the term is automatically infered
by the OCaml's type system, which helps debugging.

Running the example \verb|res| with our optimized translation,
denoted by \verb|Ex2(SPL_opt)| in the following code,
we get the correct answer for the input \verb|[1;2;3]| as follows:
\begin{lstlisting}[style=ocaml]
# let _ = let module M = Ex2(SPL_opt) in M.res;;
- : int list list = [[1]; [1; 2]; [1; 2; 3]]
\end{lstlisting}

%% file: conclusion.tex

\section{Related Work and Conclusion} \label{sec:conclusion}

In this paper, we have proposed type-preserving translation for
embedding programs with ATM into those without.  Our translation
uses multi-prompt systems and dynamic creation of prompts
to emulate two answer types in effectful terms.  We proved type preservation
for the naive and optimized translations, and implemented them in OCaml 
(and MetaOCaml) using the tagless-final style, which
we think add further assurance for type safety.

One may wonder if the reverse translation is possible. The answer is no,
as our source calculus $\lamATM$ is strongly normalizing, while the
target $\mpsr$ is not. An open question is to identify the image
of our translation which corresponds to the source calculus.


Let us briefly summarize related work.
Rompf et al.~\cite{Rompf2009} implemented
shift and reset in Scala, that allow answer-type modification.
Their source language needs relatively heavy type annotations
to be implemented by a selective CPS transformation, and 
does not allow higher-order functions.
Masuko and Asai~\cite{Masuko2011} designed OchaCaml,
which is an extension of Caml light with shift and reset.
OchaCaml fully supports ATM at the cost of redesigning the whole type system
and an extension of the run-time system.
Wadler~\cite{Wadler1994} studied monad-like structures to express
shift and reset with Danvy and Filinski's type system~\cite{Danvy1989}.
Inspired by his work,
Atkey~\cite{Atkey2006,Atkey2009}
proposed \textit{parameterised monads} as a generalization of monads.
They take two additional type parameters to express
inputs and outputs, and therefore, can express answer-type modification.
He studied categorical foundation of parameterised monads.
Kiselyov~\cite{Kiselyov2006param}
independently studied a similar notion,
and gave an implementation and programming examples.

For future work, we plan to formally prove the semantics-preservation property
mentioned in this paper.  Investigating other delimited-control operators
such as shift0/reset0 and control/prompt with answer-type modification would
be also interesting.

